\documentclass[aps,prb,twocolumn,groupedaddress,showpacs]{revtex4}
\usepackage{graphics}
\usepackage{graphicx}
\usepackage{epsfig}
\usepackage{amssymb}
\usepackage{amsmath}
\DeclareGraphicsExtensions{.jpg}

\begin{document}
\title{Transport properties of  graphene quantum dots}
\author{J. W. Gonz\'alez}
\affiliation{Departamento de F\'{i}sica, Universidad  T\'{e}cnica Federico Santa
Mar\'{\i}a, Casilla 110 V, Valpara\'{i}so, Chile}
\author{L. Rosales\cite{email}}
\affiliation{Departamento de F\'{i}sica, Universidad  T\'{e}cnica Federico Santa
Mar\'{\i}a, Casilla 110 V, Valpara\'{i}so, Chile}
\affiliation{Instituto de F\'{i}sica, Pontificia Universidad Cat\'{o}lica
de Valpara\'{\i}so, Casilla 4059, Valpara\'{i}so, Chile}
\author{P. A. Orellana}
\affiliation{Departamento de F\'{i}sica, Universidad Cat\'{o}lica del Norte,
Casilla 1280, Antofagasta, Chile}
\author{M. Pacheco}
\affiliation{Departamento de F\'{i}sica, Universidad  T\'{e}cnica Federico Santa
Mar\'{\i}a, Casilla 110 V, Valpara\'{i}so, Chile}

\date{\today}

\begin{abstract}
In this work we present a theoretical study of transport properties
of a double crossbar junction composed by segments of graphene
ribbons with different widths forming a graphene quantum dot
structure. The systems are described by a single-band tight binding
Hamiltonian and the Green's function formalism using real space
renormalization techniques. We show calculations of the local
density of states, linear conductance and I-V characteristics. Our
results depict a resonant behavior of the conductance in the quantum
dot structures which can be controlled by changing geometrical
parameters such as the nanoribbon segments widths and relative
distance between them. By applying a gate voltage on determined
regions of the structure, it is possible to modulate the transport
response of the systems. We show that negative differential
resistance can be obtained for low values of  gate and bias
voltages applied.
\end{abstract}

% 73.63.-b Electronic transport in nanoscale materials and structures
% 61.46.-w Structure of nanoscale materials
% 73.22.-f Electronic structure of nanoscale materials: clusters, nanoparticles,
% nanotubes, and nanocrystals
%\end{frontmatter}
\maketitle

\section{Introduction} \label{Intro}

In the last few years, graphene-based systems have attracted a lot
of scientific attention. Graphene is a single layer of carbon atoms
arranged in a two dimensional hexagonal lattice. In the literature,
it has been reported several experimental techniques in order to
obtain this crystal, such as mechanical peeling or epitaxial
growth\cite{Novoselov,berger1,berger2}. On the other hand, graphene
nanoribbons (GNRs) are stripes of graphene which can be obtained by
different methods like high-resolution lithography \cite{chinos},
controlled cutting processes\cite{Ci} or by unzipping multiwalled
carbon nanotubes \cite{Kosynkin}. Different graphene
heterostructures based on patterned GNRs have been proposed and
constructed, such as graphene junctions \cite{jarillo}, graphene
flakes \cite{ponomarenko}, graphene antidots superlattices
\cite{pedersen}, and graphene nano-constrictions \cite{jarillo2}.
The electronic and transport properties of these nanostructures are
strongly dependent of their geometric confinement, allowing the
possibility to observe quantum phenomena like quantum interference
effects, resonant tunneling and localization. In this sense, the
controlled modification of these quantum effects by means of
external potentials which change the electronic confinement, could
be used to develop new technological applications such as
graphene-based composite materials \cite{stankovic}, molecular
sensor devices \cite{schedin, Rosales} and  nanotransistors
\cite{Stampfer}.

\begin{figure}[ht]
\centering
\includegraphics[width=0.47\textwidth,angle=0,clip] {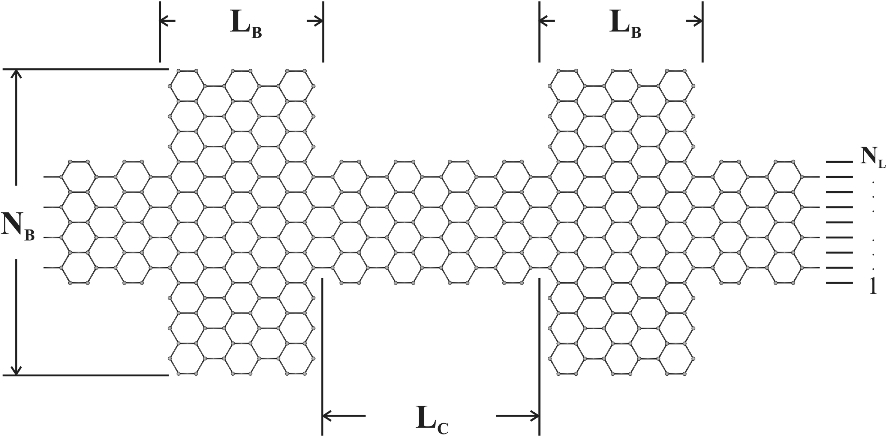}
\caption{Schematic view of a GQD structure based on leads of width
$N_{L}=9$, and a conductor region composed by two symmetrical
junctions of width $N_{B} = 21$ and length $L_{B} = 3$ separated by
a central structure of length $L_{C} = 4$ and width $N_{C} = 9$}
\label{fig:esq}
\end{figure}

In this work we study the transport properties of quantum-dot like
structures, formed by segments of graphene ribbons with different
widths connected between each other, forming a double crossbar
junction\cite{Gonzalez, chinos2}. These graphene quantum dots (GQDs)
could be versatile experimental systems which allow a range of
operational regimes from conventional single-electron detectors to
ballistic transport.  The  systems we have considered are
conductors formed by two symmetric crossbar junctions of widths
$N_{B}$ and length $L_{B}$, and a central region that separates the
junctions, of width $N_{C}$ and length $L_{C}$.  Two semi-infinite
leads of width $N_{L} = N_{C}$ are connected to the ends of the
central conductor. A schematic view of the considered system is
presented in figure \ref{fig:esq}. We  studied the different electronic states
appearing in the system as a function of the geometrical parameters
of the GQD structure. We found that the GQD local density of states
(LDOS) as a function of  the energy shows the presence of a variety
of sharp peaks corresponding to localized states  and also states
that contributes to the electronic transmission  which are
manifested as resonances in the  linear conductance. By changing the
geometrical parameters of the structure, it is possible to control
the number and  position  of these  resonances as a function of the
Fermi energy. On the other hand a gate voltage applied at selected
regions of the conductor allows  the modulation of their transport
properties exhibiting a negative differential conductance (NDC) at
low values of  the bias voltage.

\section{Model} \label{teorico}
All considered systems have been described by using a single
$\pi$-band tight binding hamiltonian, taking into account  nearest
neighbor interactions with a hopping parameter $\gamma_0 =
2.75\,eV$.  Besides, we have considered hydrogen passivation  by
setting a different hopping parameter for the carbon dimers at the
ribbons edges\cite{Son}, $\gamma _{edge} = 1.12\gamma_0$ .

The electronic properties of the systems have been calculated using
the surface Green's functions matching formalism
(SGFMF)\cite{Nardelli, Rosales}. In this scheme, we  divide
the heterostructure in three parts, two leads composed by
semi-infinite pristine GNRs, and the conductor region composed by
the double GNRs crossbar junctions, as it is shown in figure
\ref{fig:esq}.

In the linear response approach, the electronic conductance is
calculated by the Landauer formula. In terms of the conductor
Green's functions, it can be written as\cite{Datta}:
\begin{equation}\label{LandauerG}
G = \frac{{2e^2 }}{h}\bar T\left( {E } \right) = \frac{{2e^2 }}{h}
{\mathop{\rm Tr}\nolimits} \left[ {\Gamma _L G_C^R \Gamma _R G_C^A }
\right],
\end{equation}
where  $\bar T\left( {E } \right)$, is the transmission function of
an electron crossing the conductor region, $\Gamma_{L/R}=i[ {\Sigma
_{L/R} - \Sigma _{L/R} ^{\dag} }]$ is the coupling between the
conductor and the respective leads, given in terms of the
self-energy of each lead: $\Sigma _{L/R}  =
V_{C,L/R}\,g_{L/R}\,V_{L/R,C}$. Here, $V_{C,L/R}$ are the coupling
matrix elements and $g_{L/R}$ is the surface Green's function of the
corresponding lead \cite{Rosales}. The retarded (advanced) conductor
Green's function are determined by\cite{Datta}:
\begin{equation}
G_{C}^{R,A}=[E - H_C- \Sigma_{L}^{R,A}-\Sigma_{R}^{R,A}]^{-1}
\end{equation}
where $H_C$ is the hamiltonian of the conductor. In order to
calculate the differential conductance of the systems, we
determine the I-V characteristics by using the Landauer
formalism\cite{Datta}. At zero temperature, it reads
\begin{equation}\label{current}
I \left(V\right) = \frac{2e}{h}\int\limits_{\mu_0-V/2
}^{\mu_0+V/2}\,\bar T\left(E,V\right)\;dE,
\end{equation}
where $\mu_0$ is the chemical potential of the system in equilibrium
and $\bar T\left( {E,V} \right)$ is defined by equation
(\ref{LandauerG}). The  Green's functions and the coupling terms
depend on the energy and the bias voltage. We  consider  a linear
voltage drop along the longitudinal direction of the conductor and
the gate voltage is included in the on-site energy at the regions
in which this potential is applied. In what follows the Fermi energy
is taken as the zero energy level, all energies are written in terms
of the hopping parameter $\gamma_0$ and the conductance is written
in units of the quantum of conductance $G_0 = {2e^2 }/{h}$.

\section{Results and discussion} \label{Resultados}

\begin{figure}[h]
\centering
\includegraphics [width=0.47\textwidth,angle=0,clip] {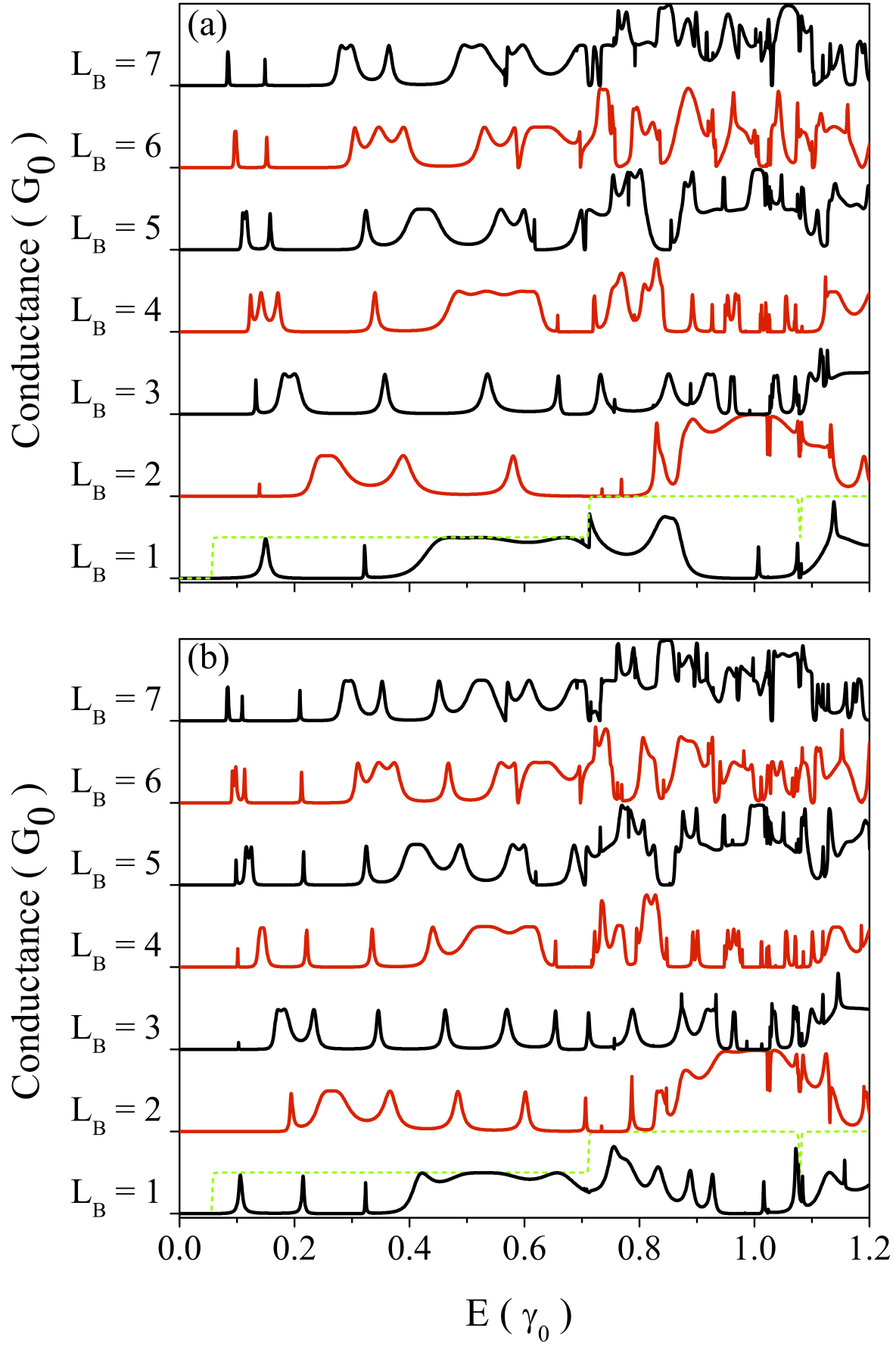}
\caption {Conductance as a function of the Fermi energy for  a graphene
quantum dot structure composed by two armchair ribbons leads of
width $N_L =5$, a double symmetric crossbar junctions of width $N_B
= 17$ and variable length, from $L_B = 1$ up to $L_B =7$. The
central region has a width $N_C =5$ and two separations (a) $L_C =
5$ and (b) $L_C = 10$.  Light green dotted line corresponds to the
conductance of a pristine $N_L= 5$  ribbon. All curves have been
shifted  $2G_0$ for a better visualization.}\label{fig:Lb}
\end{figure}

In figure \ref{fig:Lb}, we  display results of the linear conductance for a graphene quantum dot
structure formed by two armchair ribbons leads of width $N_{L}=5$
and a conductor region composed by two symmetric crossbar junctions
of width $N_{B}=17$ and variable lengths $L_{B}$ (from $1$ up to
$7$) . Two relative distances between the junctions: panel (a) $L_C = 5$ and
panel (b) $L_C = 10$ are considered and  the  conductance of a pristine $N_L= 5$
armchair nanoribbon is included as a comparison (light green
dotted line).

In both panels it is possible to observe a series of peaks at
defined energies in the conductance curves. This resonant behavior
of the electronic conductance arises from the interference of the
electronic wave functions inside the structure, which travel forth
and back forming stationary states in the conductor region
(well-like states). In order to understand these results, it is
convenient to define two energy regions, the low-energy range from
$0$ up to $0.7$ $\gamma_0$ (corresponding to the first quantum of
conductance for the pristine $N=5$ armchair ribbon) and the
high-energy range, from $0.7$ to $1.2$ $\gamma_0$ (corresponding to
the second step of conductance of the $N=5$ pristine system).

\begin{figure}[h!]
\centering
\includegraphics[width=0.47\textwidth,angle=0,clip] {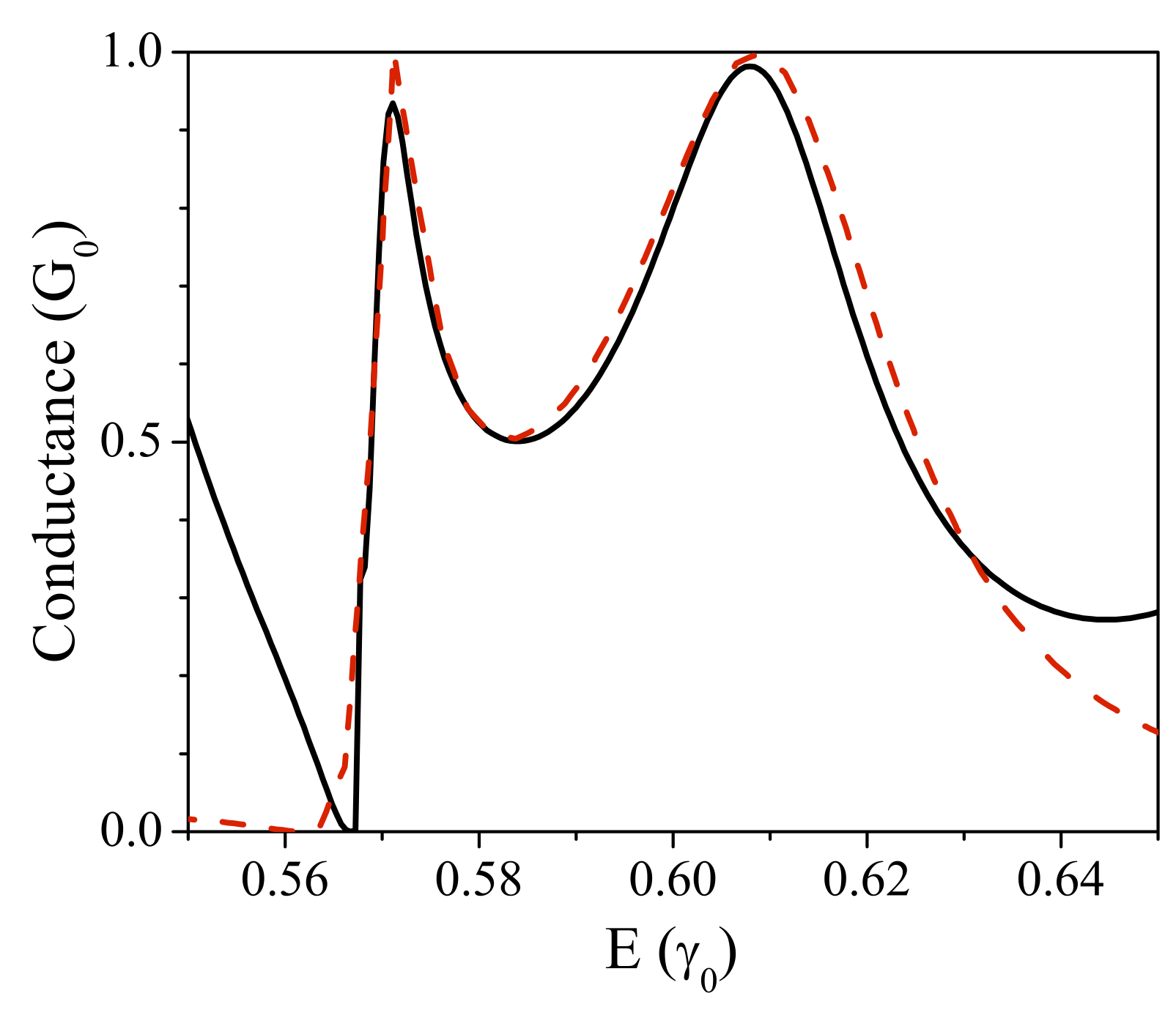}
\caption{(Color online)  Numerical adjustment of a convolution of a Fano and a  Breit -Wigner line shapes
(red dash line) and one of the conductance resonance in the
system (black solid line) with $\eta=0.01\gamma_0$, $\omega_c=0.59\gamma_0$,
x=0.7,$\tau=0.02\gamma_0$.} \label{ajuste}
\end{figure}

In the low-energy range, it is clear that the conductance peaks
correspond to resonant states belonging to the central region of the
conductor. By increasing the relative distance $L_{C}$ of the
central part of the system, the
number of allowed well-like states also increases and, as a
consequence, the conductance curves exhibit more resonances\cite{Rosales,
Gonzalez, Zhang}. The well-like states remain almost invariant under
geometrical modifications of the crossbar junctions. However, for certain
energy ranges and for particular junction lengths, the electronic transmission of the systems.
exhibits an almost constant value.
For instance, in both panels of figure \ref{fig:Lb}, for the cases of
$L_{B}=1$ and $L_{B}=4$ at the energy range 0.4 to 0.65 $\gamma_0$.
This effect corresponds to a constructive interference between well-like states
from the central region with states belonging to the crossbar
junctions regions. The different interference effects  will be clarified by
analyzing  the LDOS of these systems, which is shown next in this
paper. In the high-energy region, the conductance curves exhibit a
complex behavior as a function of the geometrical parameters of the
GQD structures. There is not a predictable behavior of the
conductance as the width and length of the crossbar junction are
increased.

It is important to pointed out, from the analysis of figure
\ref{fig:Lb}, that it is possible to identify some interesting
effects associated to well known quantum phenomena. For instance,
in panels (a) and (b) of figure \ref{fig:Lb}, for the cases $L_{B}
= 5, 6$ and $7$ at energies around $E = 0.5 \gamma_0$, it is
possible to observe a non-symmetric line shape, which corresponds
to a convolution of a Fano-like\cite{fano} and a
Breit-Wigner\cite{BreitWigner} resonances. This kind of
line-shape, has been observed before in other mesoscopic systems
by \emph{Orellana and co-workers}\cite{orellanaFBW}. In that reference,  a simple model of
two localized states with the same energy $\omega_c$ are directly coupled between each other  by a
coupling $\tau$ and indirectly coupled  throughout a common continuum. The corresponding resonances
have been adjusted by using the following expression:
\begin{equation}
T(\omega) = \frac{4\eta^2[(\omega-\omega_c)x-\tau]^{2}}{[(1-
x^{2})\eta^2-(\omega-\omega_c)^{2} +\tau^{2}]^{2} +
4(\omega-\omega_c -\tau x)^{2}} \label{fit}
\end{equation}

\noindent
where $\eta$ is the width of a localized state coupled
to the continuum  and $x$ defines the degree of asymmetry of the
system.

We realize that a possible interference mechanism occurring  in our considered system can be explained with
the above model, which helps to get an intuitive
understanding of the origin of some conductance lines
shape. In figure \ref{ajuste} we have plotted a particular conductance resonance
and the corresponding fitting  given by the model
represented by equation (\ref{fit}), where it is observed the
good agreement between both curves.

\begin{figure}[h!]
\centering
\includegraphics[width=0.47\textwidth,angle=0,clip] {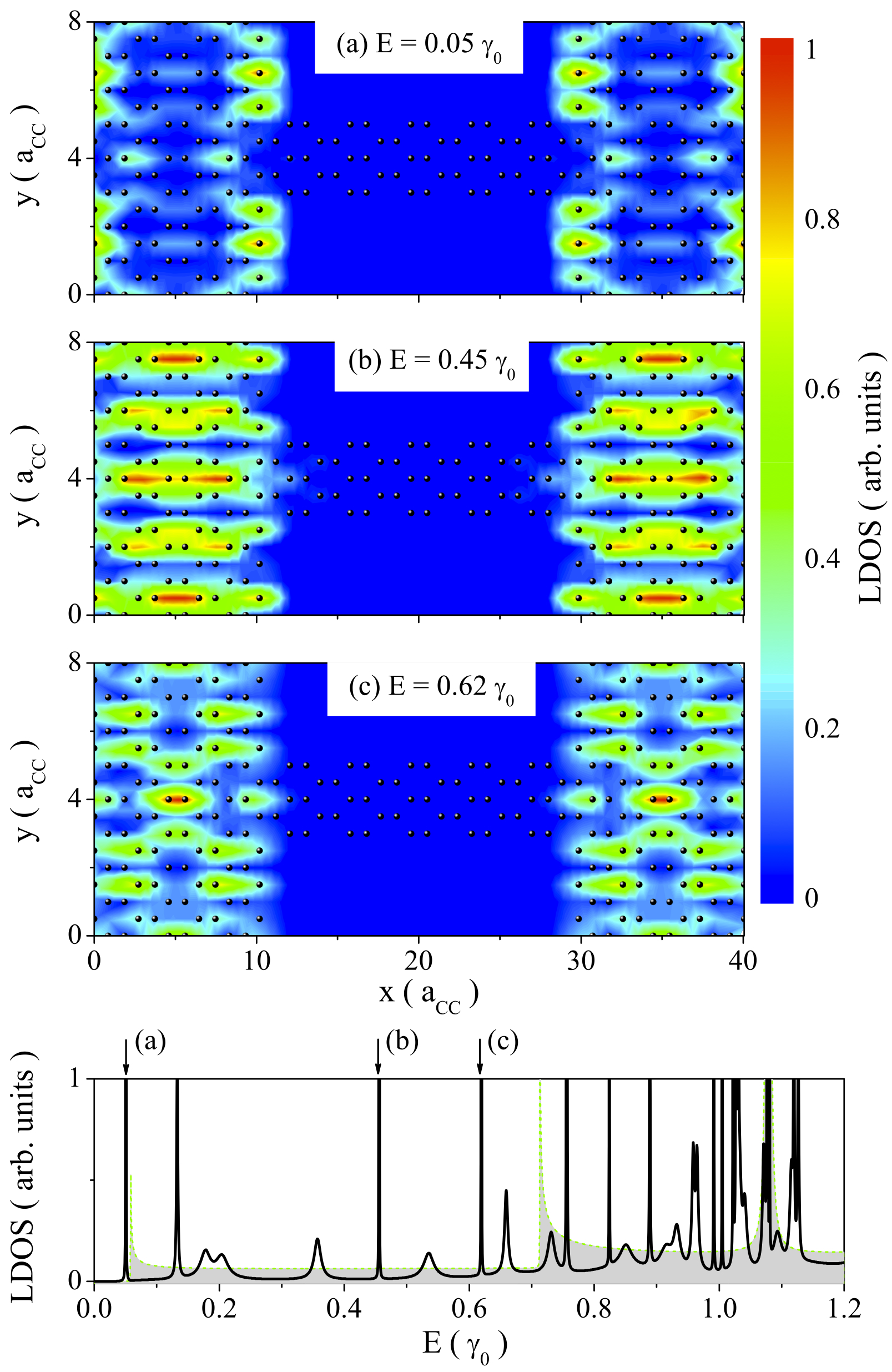}
\caption{LDOS for a GQD formed by a double crossbar junction of
width $N_{B}$= 17 and length $L_{B}$= 3 separated by a central
region of width $N_{C}$= 5 and length $L_{C}$=5. Panel (a), (b) and
(c) correspond to the contour plots of some sharp LDOS resonances
marked in the bottom plot. As a reference, the LDOS of a pristine
N=5 armchair ribbon is plotted as a dotted green
line} \label{fig:odos_barrera}
\end{figure}

In what follows we focus our analysis on the resonant behavior exhibited by
the conductance curves, analyzing the different electronic states in the conductor.
We have performed calculations of the
spatial distribution of LDOS for certain energies corresponding to
different states present in the conductor. In the bottom panel of
figure \ref{fig:odos_barrera} we show results for the  LDOS as a
function of the Fermi energy, for a GQD structure formed by a double
crossbar junction of width $N_{B}=17$ and length $L_{B}=3$,
separated by a central region of width $N_{C}=5$ and length
$L_{C}=5$.

\begin{figure}[h!]
\centering
\includegraphics[width=0.47\textwidth,angle=0,clip] {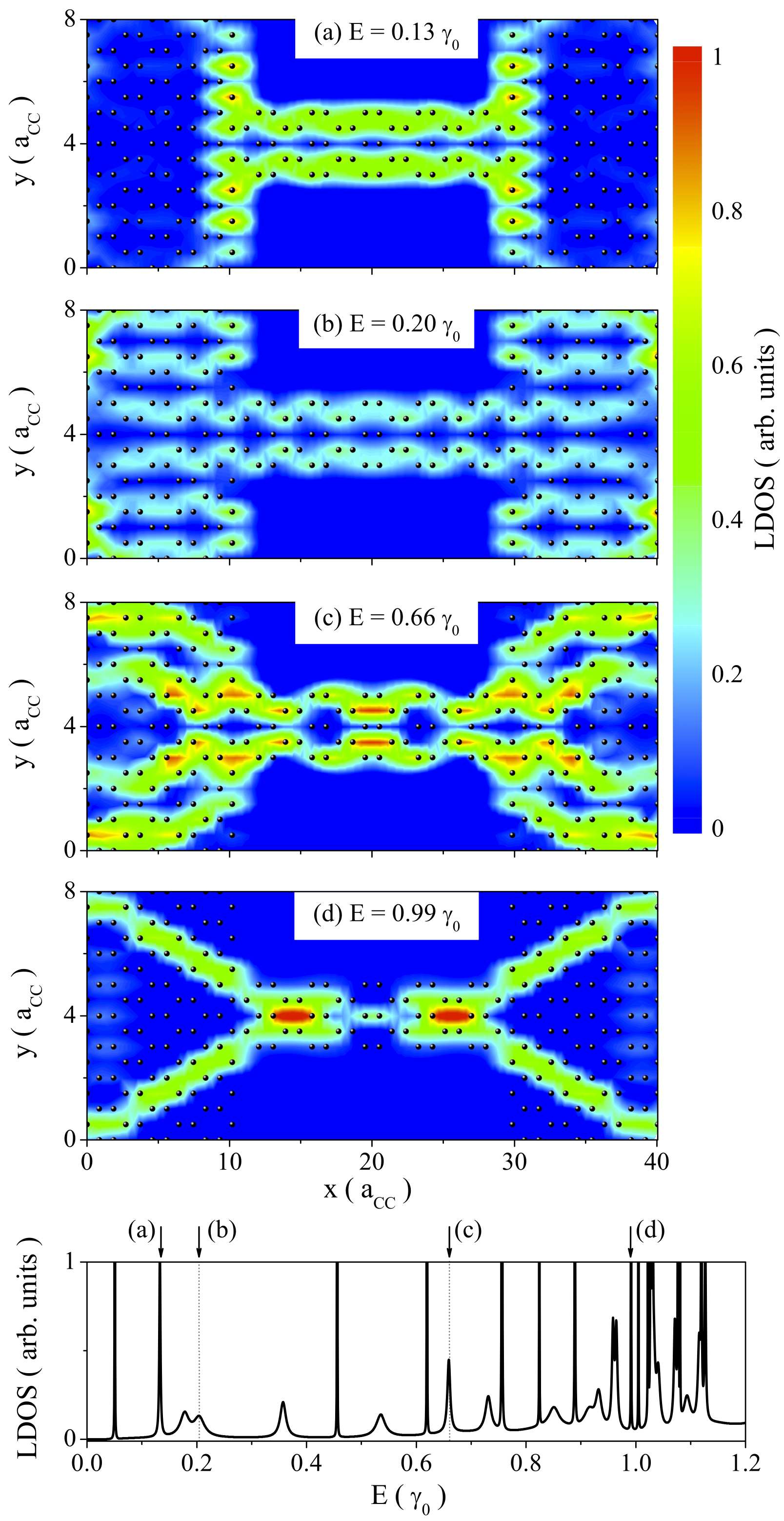}
\caption{LDOS for a GQD structure formed by a double crossbar
junction of width $N_{B}$= 17 and length $L_{B}$= 3 separated by a
central region of width $N_{C}$= 5 and length $L_{C}$=5. Panel
(a),(b), (c) and (d) correspond to the contour plots of those
resonant states marked in the LDOS  plot displayed  at the bottom.}
\label{fig:odos_barrera2}
\end{figure}

We start our analysis focusing on the some sharp states present in the
curve LDOS vs energy of this figure. We have marked the first three
sharp states in this LDOS plot with the letters (a), (b) and (c) and
we have calculated the spatial distribution of these states,
representing by the corresponding contour plots exhibited in the
figure. These states are completely localized at the crossbar
junctions, and as we establish in a previous work\cite{gonzalez2},
their correspond to bound states in the continuum
(BICs)\cite{wigner, chinos1,pet1}. It is not expected that these kind of
states play a role in the electronic transport of these  GQD
structures, which is shown at the corresponding conductance curves
of figure \ref{fig:Lb}.

\begin{figure*}[ht]
\centering
\includegraphics[width=0.85\textwidth,angle=0,clip]{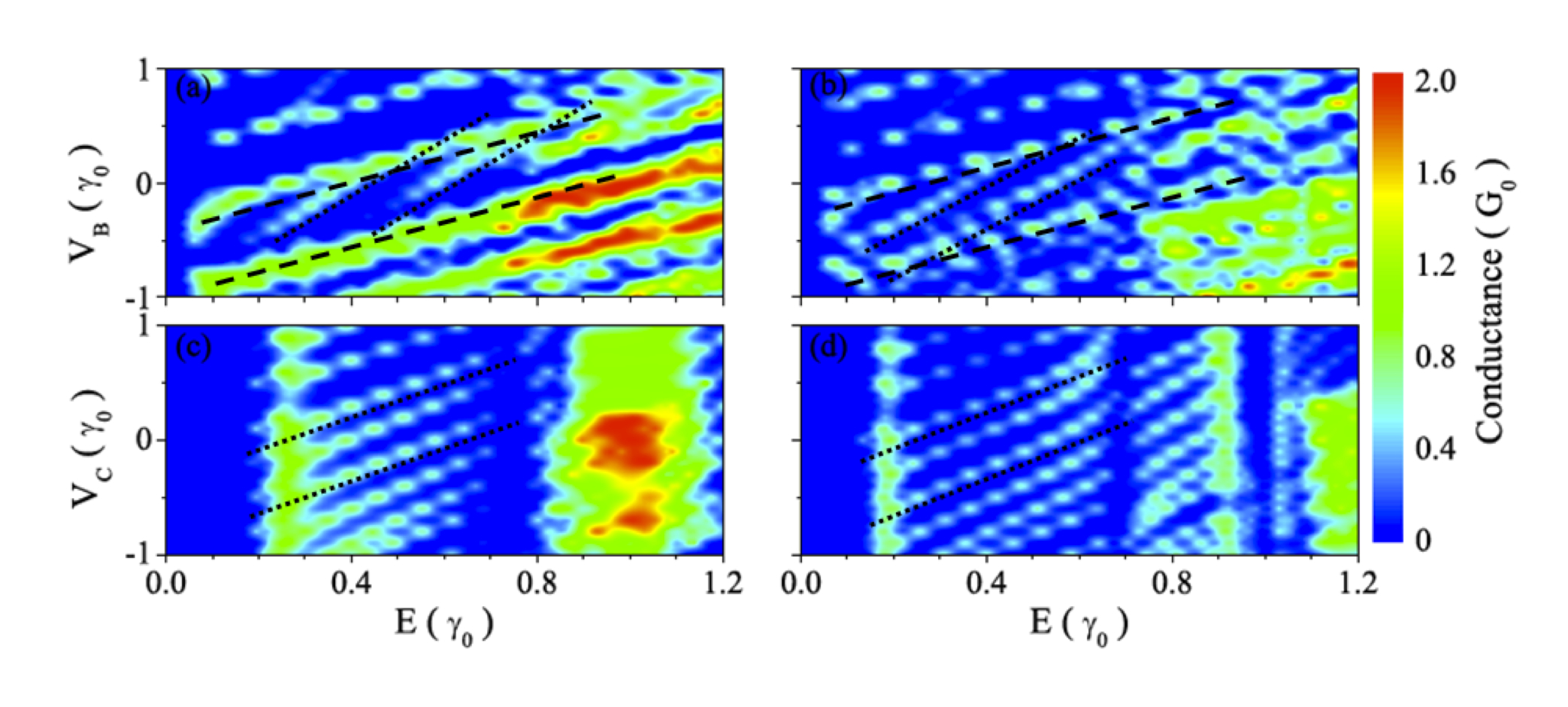}
\caption{Conductance as a function of Fermi energy and gate voltage
for GQDs composed by leads of width $N_L = 5$, two crossbar
junctions of width $N_B = 17$, a central part of width $N_C = 5$ and
length $L_C = 5$.  Panels (a) and (c) correspond to junctions of
length $L_B = 2$, while panels (b) and (d) correspond to junctions
of length $L_B = 3$. In the upper panels the gate voltage is applied
at the crossbar  regions, and  in the lower panels the gate voltage
is applied at the central structure. The black segmented lines highlight
different slopes discussed in the text}\label{fig:cp_e_vg_cond}
\end{figure*}

In what follows we will focus our analysis to those states that
contribute to the conductance of the systems. In figure
\ref{fig:odos_barrera2} it is shown the spatial distribution of LDOS
for a GQD formed by a double crossbar junction of width $N_{B}=17$
and length $L_{B}=3$, separated by a central region of width
$N_{C}=5$ and length $L_{C}=5$. As a reference, at the bottom panel
we have included a plot with the LDOS versus Fermi energy  of the
GQD system, there we have marked with letters (a), (b), (c) and (d)
four particular energy states. The corresponding contour plots are
displayed at the upper part of the figure.

In these plots, it is possible to observe the resonant behavior of
these states, which are completely distributed along the GQD
structure, presenting a maximum  of  the  probability density  at
the center of the system. This condition favors the alignment of the
electronic states of the leads with the resonant states in the
conductor and consequently, a unitary transmission at those energy
values is expected. This behavior is reflected as a series of
resonant peaks   in the conductance of the system that could be
controlled by means of  the geometrical parameters of the GQD, as it
is shown in figure \ref{fig:Lb}. At  higher energies there is an
interplay between localized states in the crossbar junctions and
resonant states in the central region of the conductor. As it is
shown in panel (d) of figure \ref{fig:odos_barrera2}, some states
are strongly dependent of the geometry of the junctions, therefore
for some particular configuration it is possible to observe a
non-null transmission at these energies, while for other cases,
there are destructive interference mechanisms that suppress
completely the transmission at that energy region.

We have studied different
configurations of GQDs structures, varying systematically some
geometric parameters. We have observed a quite
general behavior of such resonant conductors with the presence of
sharp and resonant states. Depending of each particular system, it can be observed changes in the
number and position in energy of these states, as well as in their spatial distribution.
The different intensity of the peaks in the
LDOS curves depends on the spatial distribution of the states.
There are states completely extended along the conductor (like in
panel (b) of figure \ref{fig:odos_barrera2} ) which generate
wider and less intense peaks. On the other hand, there are
resonant states which are more concentrated in certain regions of the
conductor (like as panel (c) and (d) of figure
\ref{fig:odos_barrera2}) which generate sharper and more intense
peaks in the LDOS.

Now we focus our analysis on the effects of an applied
gate voltage on the transport properties of GQD structures.  Results
of the conductance as a function of the Fermi energy, for different
values of a gate voltage applied in selected regions of a GQD are
shown in figure \ref{fig:cp_e_vg_cond}. The systems are composed by
 leads of width $N_L = 5$, two crossbar
junctions of width $N_B = 17$, a central part of width $N_C = 5$ and
length $L_C = 5$. Panels (a) and (c) correspond to junctions of
length $L_B = 2$, while panels (b) and (d) correspond to junctions
of length $L_B = 3$.  Finally, upper panels correspond to a gate
voltage applied at the crossbar junctions regions, whereas the lower
panels correspond to a gate voltage applied at the central part of
the structure.

In these contour plots of conductance, it is possible to observe the
behavior of the resonant states of the system with a gate voltage
applied at different regions of the structure. The lower panels of
the figure \ref{fig:cp_e_vg_cond} show the case of a gate voltage
applied at the central region of the considered GQDs.  In these plots the  linear dependence of the  conductance
resonances as a function of the gate voltage is manifested. It can be shown that
the electronic states of a pristine armchair graphene ribbon are regularly spaced in the whole energy
range\cite{nakada,wakabayashi}, therefore, as the gate voltage is
applied, there will be a high
probability that  the lead states become  aligned with
the resonant states in  the central region of the structure.  This
behavior in completely general and independent of the width $L_{B}$
of the crossbar junctions. The linear behavior of the conductance
peaks could be useful in nanoelectronics devices, due to the
possibility of control the current flow through these systems. This
argument will be more clear with the analysis of the differential
conductance, which is shown below in this paper.

The case of a gate voltage applied at the crossbar junction
regions is shown in the upper panel of figure
\ref{fig:cp_e_vg_cond}. The conductance behavior in more complicated
to analyze, nevertheless it is still possible to observe a linear
dependence of the resonant states with the gate voltage. However,
two different slopes can be noticed, for states belonging to the crossbar junctions (lower slope)
and for states belonging to the central region of the conductor
(higher slope).
Besides, the panels exhibit  an
important reduction in  the conductance gap, for different values of
gate voltage. This effect is mainly produced by an energy shift of
the localized states at the junctions, which induces a less
destructive interference with the resonant states. It is important
to point out that  this effect only can be observed  because the
gate potential is applied simultaneously to both crossbar junctions,
otherwise, the conductance gap would be not noticeably affected by
the gate potential.
Finally, the dark (blue online) regions present in figure \ref{fig:cp_e_vg_cond},
occur at energy ranges around  the LDOS singularities of the
pristine N=5 AGNR. At these energies, it appear the second and the third
allowed states appear, which interrupt  the linear behavior  of the conductance resonances as
a function of the gate voltage.

In order to understand the presence of two different slopes in the upper panel of
figure \ref{fig:cp_e_vg_cond}, we present a simple model which keeps the underlying physics of the
considered system and allow us to explain  qualitatively our results.

\begin{figure}[ht]
\centering
\includegraphics[width=0.47\textwidth,angle=0,clip] {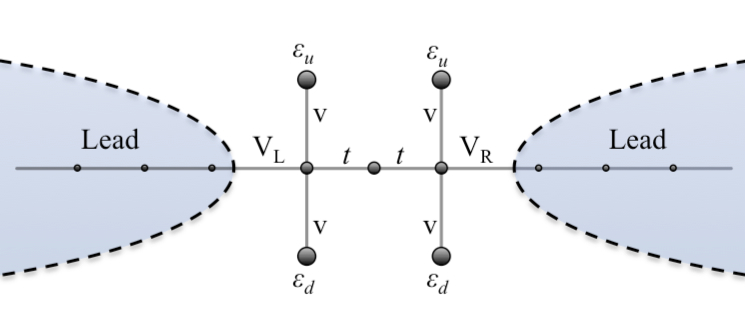}
\caption{A simple model of two crossbar junctions formed by two
quantum dots, coupled to a linear chain of sites.} \label{modelo}
\end{figure}

The scheme  showed in figure \ref{modelo} is a simple representation of our conductor. The system is composed by a
linear chain of three sites, which are connected to two
semi-infinite leads. We have considered four quantum dots connected
to the extremes of the chain forming a double crossbar junction, at
which we have applied symmetrically a gate voltage $V_g$. This
potential will modify the on-site energy of the dots by a linear
shift of energy proportional to the gate voltage amplitude.

By using Dyson equation, it is possible to calculate the Green's
function of the central site of the chain labeled by $0$, which
takes the form:
\begin{equation}
G_{00} = \frac{1}{\omega - \varepsilon_{0} - \Sigma}\label{mod1}
\end{equation}

\noindent where $\omega$ is the energy of the incident electrons,
$\varepsilon_{0}$ is the central on-site energy and the self energy
$\Sigma$ is given by the following expression (see Appendix for a
detailed deduction):

\begin{equation}
\Sigma = \frac{v^{2}}{\left(\omega -
\frac{v^{2}}{\omega-V_g}\right)^{2}+\widetilde{\Gamma}^{2}}\left[\left(\omega
- \frac{v^{2}}{\omega-V_g}\right)+ i\widetilde{\Gamma}\right]
\label{mod2}
\end{equation}

In this model, the self energy of the Green's functions of the
central region acquires a real part that depends on the gate voltage
applied to the crossbar junction region. As a consequence, two
different slopes appear in the behavior of the conductance peaks as
a function of the gate voltage. One of these slopes corresponds to
the direct evolution of the states belonging to the crossbar
junctions as a function of the gate voltage (lower slope), and the
higher slope corresponds to the indirect states belonging to the
central region.

\begin{figure}[ht]
\centering
\includegraphics[width=0.47\textwidth,angle=0,clip] {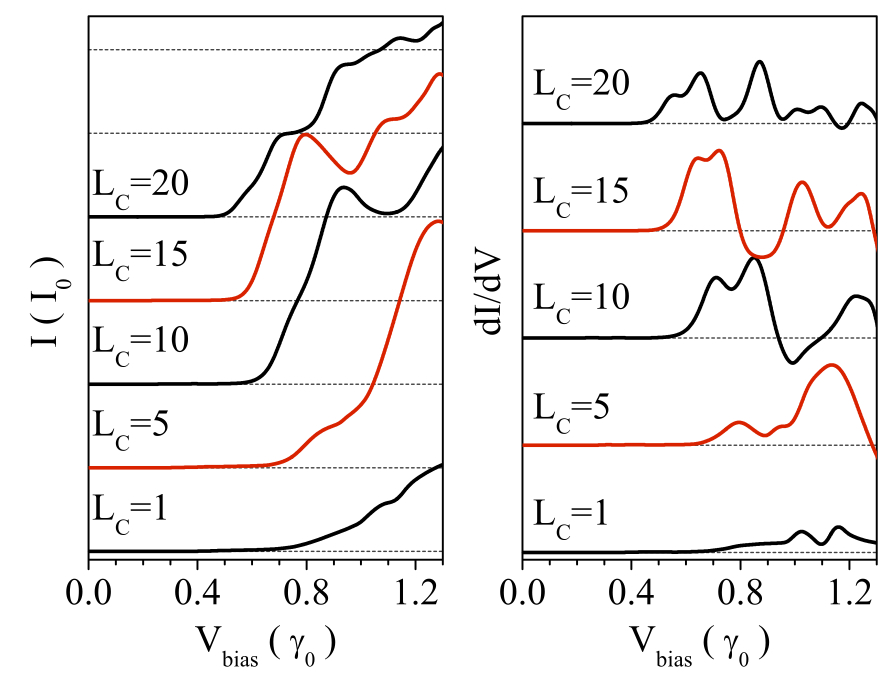}
\caption{(a) Current versus bias voltage and (b) Differential
conductance as a function of bias voltage for a GQD composed by two
crossbar junctions of length $L_{B} = 2$ and width $N_{B} = 17$
separated by a central region of width $N_{C} = 5$ and a variable
length from $L_{C} = 1$ up to $L_{C} = 20$. All curves have been shifted
for a better visualization.} \label{corr}
\end{figure}

Now we focus our analysis on the I-V characteristics and the
differential conductance of these resonant GQDs.  Figure \ref{corr}
shows results of these transport properties for a conductor formed
by two crossbar junctions of width $N_{B} = 17$, length $L_{B} = 2$
and a central region of width $N_{C} = 5$. In this plots  the length
of the central structure is varied from $L_{C} = 1$ up to $L_{C} =
20$. In panel (a) it  is possible to observe that for a very small
separation between both  junctions, the I-V characteristics exhibit
abrupt slope changes, and oscillations for certain ranges of the
bias voltage. This behavior is produced by the increasing number of
resonant  states  as a result of  the enlargement of the conductor
central region. The applied bias voltage allows the continuum
alignment of the resonant states of the system with the electronic
states of the leads, leading to variations of the current intensity.
On the other hand, every I-V curve shows a wide  gap of zero current
until certain bias  voltage. This threshold value exhibits a linear
dependence of  the length of the central region of the conductor. As
the relative distance between the junctions is increased, there are
more resonant states  available at lower energies because the
electronic confinement in this region is weaker, therefore, the
electronic transmission under a bias voltage is possible at lower
voltage values.

The abrupt changes of  the current as a function of the bias voltage
are clearly reflected in  the differential conductance of these
systems. In panel (b) of figure \ref{corr} it is possible to observe
the behavior  of the $dI/dV - V$ curves as a function of the length
of the central region of the conductor. As this region is enlarged,
the oscillations in the differential conductance become more
evident. Each time the bias voltage aligns the resonant states of
the conductor with leads states, the current will increase and  a
positive change in the differential conductance occurs. However, if
the bias voltage is not enough  to align the states, the current
drops and the differential conductance becomes negative in a range of voltage. This can be seen in the cases  $L_C = 10$,
$L_C =15$ and $L_C =20$ in the figure \ref{corr} (b). The bias
voltage value at which occurs negative differential conductance
(NDC)\cite{NDC1,NDC2,NDC3} depends directly on the relative distance
between the crossbar junctions.

\begin{figure}[ht]
\centering
\includegraphics[width=0.47\textwidth,angle=0,clip] {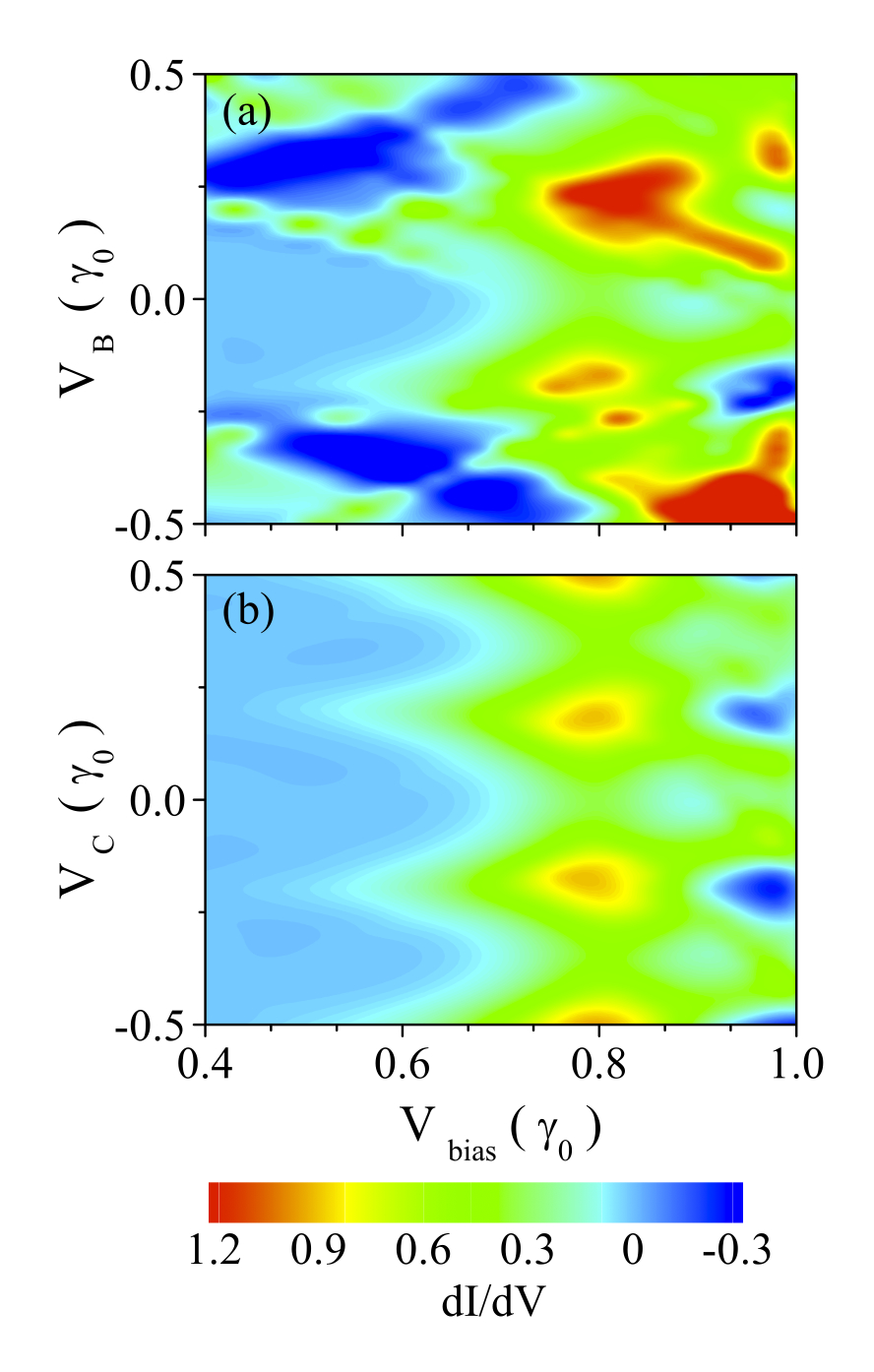}
\caption{Differential conductance of a GQD structure composed by two
crossbar junctions of length $L_{B} = 2$ and width $N_{B} = 17$
separated by a central region of width $N_{C} = 5$ and length $L_{C}
= 5$. The gate voltage is applied at selected regions of the
conductor: (a) on the crossbar junction region and (b) on the
central region.} \label{fig:cp_e_vg_dI}
\end{figure}

Now we focus our analysis  in the effects  of a gate voltage  on the
differential conductance of the systems.  In figure
\ref{fig:cp_e_vg_dI} we present results of the differential
conductance of a GQD composed by two crossbar junctions of length
$L_{B} = 2$ and width $N_{B} = 17$ separated by a central region of
width $N_{C} = 5$ and length $L_{C} = 5$. The gate potential has
been applied at selected regions of the structure: (a) crossbar
junction regions and (b) at the central region.  In panel (b) of
this figure  it can be observed a periodic modulation of the
differential conductance as a function of the gate voltage applied
at the central region of the GQD. This behavior is determined and
directly related with the linear evolution of the resonant states of
the conductor, and consequently, with the peaks of conductance of
the system [panel (c) and (d) of figure \ref{fig:cp_e_vg_cond}]. For
instance, in the configuration considered in this figure, it can be
noted abrupt changes of the differential conductance as a function
of the gate voltage  for  bias voltages values around $0.8\gamma_0$,
which indicates an abrupt increasing of the current flowing through
the conductor. This behavior is very general for other systems
studied, which suggests possible applications in the development of
GQDs based electronic devices.

In the case in which the  gate voltage is applied simultaneously at
the crossbar junction regions, (panel (a) of this figure)  it is
still possible to note certain regularity in the dependence of
differential conductance as a function of the external potentials,
although  a periodic modulation  is not clearly observed. However,
in this configuration of applied potentials our results show that
the GQD structure exhibits NDC at very low values of bias voltage
and gate voltages. This behavior can be understood by observing
panels (a) and (b) of figure \ref{fig:cp_e_vg_cond}, where the
contour plots show  areas where the conductance is completely
suppressed at low energies, for different values of gate voltage. We
have observed this kind of behavior in every configuration
considered in this work.

In relation to the practical limitations of our calculations we would like to mention that despite of
the size of the structures used  in this work are below the limit  of the actual  experimentally realizable,
our calculations can be scaled to structures of bigger sizes. By other hand  our model does not include disorder or electron-electron interaction,
nevertheless we are convinced that our results will be robust under these kind of effects as it is in mesoscopic systems.
For instance it is known that in quantum dots, the resonant tunneling and the Fano effect survive
to the effect of the electron-electron interaction\cite{Kouwenhoven,kobayashi}.

\section{Summary} \label{Summary}

In this work we have analyzed the transport properties of a GQD
structure formed by a double crossbar junction made of segments of
GNRs of different widths. We have focused our analysis on the
dependence of the electronic and transport properties with  the
geometrical parameters of the system looking for the modulation of
these properties through  external potentials applied to the
structure. Our results depict a resonant behavior of the conductance
in the  quantum dot structures  which can be controlled by changing
geometrical parameters such as  nanoribbon widths and
relative distance between them. We  have explained our results in
terms of the analysis of the different electronic states of  the system.
The possibility of modulating  the transport response
by applying a gate voltage on determined regions of the structure
has been explored and  it has been found that negative differential conductance   can be obtained
for low values of  the   gate and bias applied voltages. Our results
suggest that  possible applications with  GQDs can be developed for
new electronic devices.

\section*{Acknowledgments}
The authors acknowledge the financial support of USM
110971 internal grant, FONDECYT program grants 11090212, 1100560 and
1100672. LR also acknowledges to PUCV-DII grant 123.707/2010.

\section{APPENDIX: Green's function of the simple model}

In section \ref{Resultados}, we have introduced a simple model in
order to explain the different slopes of figure
\ref{fig:cp_e_vg_cond}. This model is composed by a linear chain of
three sites, of the same energy, at which we have coupled four
quantum dots (QDs) of energies $\varepsilon_{n}$ ($n$ =
u,d), forming a crossbar junction configuration exhibited in figure
\ref{modelo}. This simple scheme is very useful to qualitatively
explain the electronic behavior  of graphene quantum dot that we have studied.

Let us start with the hamiltonian of the system described by figure
\ref{modelo}:
\begin{equation}
H_{T} =  H_{leads} + H_{c} + H_{c,leads}\label{ap1}
\end{equation}
\noindent
where the hamiltonian of the leads $H_{leads}$ is given by:
\begin{equation}
H_{leads} =
\sum_{k,\alpha(L,R)}\varepsilon_{k,\alpha}c^{\dag}_{k,\alpha}c_{k,\alpha}\label{ap2}
\end{equation}
\noindent
the conductor hamiltonian $H_{c}$ is given by;
\begin{equation}
\begin{split}
&H_{c} =\sum^{1}_{i=-1}\varepsilon_{i}f^{\dag}_{i}f_{i} \:+\:
t\sum^{1}_{i=0}\left(f^{\dag}_{i-1}f_{i} + hc \right)+ \\
&\sum_{m(-1,1)}\:\sum_{n(u,d)}\left[\varepsilon_{m,n}\;d^{\dag}_{m,n}\;d_{m,n}
+ v\left(d^{\dag}_{m,n}\;f_{m} + hc \right) \right],
\end{split} \label{ap3}
\end{equation}
\noindent
and finally the leads-conductor hamiltonian is given by:
\begin{equation}
H_{c,leads} =
\sum_{k,\alpha(L,R)}\:\sum_{m(-1,1)}\;V_{\alpha}\left(f^{\dag}_{m}\;c_{k,\alpha}+
hc\right)\label{ap4}
\end{equation}

By using the Dyson equation, it is possible to calculate the Green'
function of site $0$. Following a standard procedure we have
obtained:
\begin{eqnarray}
G_{00} &=& g_0 \; + \; g_0\,v\,G_{01}\;
+\;g_0\,v\,G_{0\bar{1}}\\\label{ap5} 
G_{10}&=&g_1\,v\,G_{00}\;+\;g_1\sum_{k}\;V_{R}G_{k_R,0}\\
G_{\bar{1}0}&=&g_0\,v\,G_{\bar{1}0}\; + \; g_{\bar{1}}\sum_{k} V_{L}
G_{k_L,0},
\end{eqnarray}
\noindent
where $G_{k_R,0}=g_k\,V_{R}G_{10}$ and
$G_{k_L,0}=g_k\,V_{L}G_{\bar{1}0}$.

Replacing these expression in the previous set of equations, we
obtain:
\begin{eqnarray}
G_{10}&=&\frac{g_1\,v\,G_{00}}{1-
g_1\;\sum_{k_R}\,V^{\dag}_{R}\;g_{k_R}\;V_R}\\
G_{\bar{1}0}&=&\frac{g_{\bar{1}}\,v\,G_{00}}{1- g_{\bar{1}}\;
\sum_{k_L}\,V^{\dag}_{L}\;g_{k_L}\;V_L}\\\nonumber\label{ap6}
\end{eqnarray}

Considering $g_{0} = 1/(\omega-\varepsilon_{0})$, and replacing the
above expressions for $G_{10}$ and $G_{\bar{1}0}$, $G_{00}$ reads:
\begin{equation}
G_{00} = \frac{1}{\omega - \varepsilon_{0} - \Sigma}\label{ap8}
\end{equation}
\noindent
where the self-energy is defined by the following expression:
\begin{equation}
\Sigma = \frac{g_{1}\,v^{2}}{1- g_{1}\;i\Gamma_R}\; + \;
\frac{g_{\bar{1}}\,v^{2}}{1- g_{\bar{1}}\;i\Gamma_L} \label{ap9}
\end{equation}
\noindent
with  $\Gamma_{R}=\sum_{k_R}V^{\dag}_{R}g_{k_R}\;V_R$ and
$\Gamma_{L}=\sum_{k_L}\,V^{\dag}_{L}g_{k_L}V_L$.

Using the expressions for the on-sites Green's functions for the
sites $\bar{1}$ and $1$ given by:
$g_{\bar{1}}=1/(\omega-\varepsilon_{\bar{1}})$ and
$g_1=1/(\omega-\varepsilon_1)$, and considering a gate voltage $V_g$
applied to the QDs, which redefine their on-sites energies by \\
$\tilde{\varepsilon}_{n}=\varepsilon_{n}+V_g$, it is
possible to write an expression for the self-energy of the systems
as:

\begin{equation}
\Sigma = \frac{v^{2}}{\left(\omega -
\frac{v^{2}}{\omega-V_g}\right)^{2}+\widetilde{\Gamma}^{2}}\left[\left(\omega
- \frac{v^{2}}{\omega-V_g}\right)+ i\widetilde{\Gamma}\right]
\label{ap11}
\end{equation}
\noindent
where we have considered a  symmetric  system ($\Gamma_L$= $\Gamma_R$). In this approach,
it is possible to write a compact form for the self-energy given
in equation (\ref{ap11}), which contains a real ( gate voltage
dependent) and an imaginary part.


\begin{thebibliography}{100}

\bibitem[*]{email} Electronic address: luis.rosalesa@usm.cl


 \bibitem{Novoselov} K. S. Novoselov, A. K. Geim, S. V. Morozov, D. Jiang, Y. Zhang,
S. V. Dubonos, I. V. Grigorieva, and A. A. Firsov, Science
\textbf{306}, 666 (2004).

\bibitem{berger1} C. Berger, Z. Song, T. Li, X. Li, A. Y. Ogbazghi, R. Feng, Z.
Dai, A. N. Marchenkov, E. H. Conrad, P. N. First, and W. A. de Heer,
J. Phys. Chem. B \textbf{108}, 19912 (2004).

\bibitem{berger2} C. Berger, Z. Song, X. Li, X. Wu, N. Brown, C. Naud, D. Mayou,
T. Li, J. Hass, A. N. Marchenkov, E. H. Conrad, P. N. First, W. A.
de Heer, Science \textbf{312}, 1191 (2006).

\bibitem{chinos} X. Li, X. Wang, L. Zhang, S. Lee, H. Dai, Science \textbf{319}, 1229 (2008).

\bibitem{Ci} L. Ci, Z. Xu, L. Wang, W. Gao, F. Ding, K.F. Kelly, B. I. Yakobson and P. M. Ajayan,
Nano Res. \textbf{1}, 116 (2008).

\bibitem{Kosynkin}  D. V. Kosynkin, A. L. Higginbotham, A. Sinitskii, J. R. Lomeda, A. Dimiev,
B. K. Price and J. M. Tour, Nature \textbf{458}, 872 (2009); M.
Terrones, Nature \textbf{458}, 845 (2009).

\bibitem{jarillo} B. Oezyilmaz, P. Jarillo-Herrero, D. Efetov, D. Abanin, L. S.
Levitov, and P. Kim, Phys. Rev. Lett. \textbf{99}, 166804 (2007).

\bibitem{ponomarenko} L. A. Ponomarenko, F. Schedin, M. I. Katsnelson, R. Yang, E. W.
Hill, K. S. Novoselov and A. Geim, Science \textbf{320}, 356 (2008)
; J. W. González, H. Santos, M. Pacheco, L. Chico, and L. Brey,
Phys. Rev. B  \textbf{81}, 195406 (2010).

\bibitem{pedersen} T. G. Pedersen, C. Flindt, J. Pedersen, N. Mortensen, A.
Jauho, K. Pedersen,  Phys. Rev. Lett. \textbf{100}, 136804 (2008).

\bibitem{jarillo2} B. Oezyilmaz, P. Jarillo-Herrero, D. Efetov, and P. Kim, Appl.
Phys. Lett. \textbf{91}, 192107 (2007)

\bibitem{stankovic} S. Stankovich, D. A. Dikin, G. H. B.
Dommett, K. M. Kohlhaas, E. J. Zimney, E. A. Stach, R. D. Piner, S.
T. Nguyen and R. S. Ruoff, Nature \textbf{442}, 282 (2006).

\bibitem{schedin} F. Schedin, A. Geim, S. Morozov, E. Hill, P. Blake, M.
Katsnelson, K. Novoselov, Nature Materials \textbf{6}, 652 (2007).

\bibitem{Rosales} L. Rosales, M. Pacheco, Z. Barticevic, A. Latg\'e, and P. Orellana,
 Nanotechnology \textbf{19}, 065402 (2008); Nanotechnology \textbf{20},
 095705, 2009.

\bibitem{Stampfer} C. Stampfer, E. Schurtenberger, F. Molitor, J. Güttinger, T. Ihn
and K. Ensslin,  Nano Letters \textbf{8}, 2378 (2008).

\bibitem{nakada} K. Nakada, M. Fujita, G. Dresselhaus and M. S.
Dresselhaus, Phys. Rev. B \textbf{54}, 17954 (1996).

\bibitem{wakabayashi} K. Wakabayashi, Phys. Rev. B \textbf{64},
125428 (2001).

\bibitem{Gonzalez} J. W. Gonz\'alez, L. Rosales, M. Pacheco,  Physica B: Cond. Matt.  \textbf{404}, 2773 (2009).

\bibitem{chinos2} B.H. Zhou, W.H. Liao, B.L. Zhou, K. Q. Chen, and G.H.
Zhou, Eur. Phys. J. B \textbf{76}, 421 2010.

\bibitem{Son}Young-Woo Son, M. L. Cohen, and S. G. Louie, Phys. Rev. Lett. \textbf{97}, 216803 (2006).

\bibitem{castro} A. H. Castro Neto, F. Guinea, N. M. R. Peres,  K. S. Novoselov,
A. K. Geim,  Rev. Mod. Phys. \textbf{81}, 109 (2009).

\bibitem{Nardelli} M. Nardelli, Phys. Rev. B \textbf{60}, 7828 (1999).

\bibitem{Datta} S. Datta, \emph{Electronic Transport properties of mersoscopic systems}(Cambridge
University Press, Cambridge, 1995).

\bibitem{Zhang} Z.Z. Zhang, Kai Chang and K.S. Chan, Nanotechnology \textbf{20}, 415203
(2009).

\bibitem{fano} U. Fano, Physical Review \textbf{124}, 1866 (1961)

\bibitem{BreitWigner} G. Breit and E. Wigner, Physical Review \textbf{49}, 519 (1936).

\bibitem{orellanaFBW} P.A. Orellana, M.L. Ladr\'{o}n de Guevara and F. Claro, Phys. Rev. B \textbf{70},
233315 (2004).

\bibitem{wigner} J. von Neumman and E. Wigner, Z. Phys. \textbf{30},
465 (1929).

\bibitem{chinos1} F. OuYang, J. Xiao, R. Guo, H. Zhang and H. Xu, Nanotechnology \textbf{20},
 055202 (2009).

\bibitem{pet1} A. Matulis and F. M. Peeters, Phys. Rev. B \textbf{ 77}, 115423 (2008).

\bibitem{gonzalez2} J. W. Gonz\'alez, M. Pacheco, L. Rosales and P. A.
Orellana, Europhysics Letters (EPL) \textbf{91}, 66001 (2010).

\bibitem{NDC1} M.Y. Han, B. Ozyilmaz, Y. Zhang, and P. Kim, Phys. Rev. Lett. \textbf{98}, 206805
(2007).

\bibitem{NDC2} V. N. Do and P. Dollfus, J. Appl. Phys. \textbf{107},063705
(2010).

\bibitem{NDC3} H. Ren, Q. Li, Y. Luo, and J. Yang, Appl. Phys. Lett. \textbf{94}, 173110
(2009).

\bibitem{Kouwenhoven} S. M. Cronenwett, T. H. Oosterkamp, and L. P. Kouwenhoven, Science \textbf{281},
540 (1998).

\bibitem{kobayashi} Masahiro Sato, Hisashi Aikawa, Kensuke Kobayashi, Shingo Katsumoto, and Yasuhiro Iye
Phys. Rev. Lett. \textbf{95}, 066801 (2005).

\end{thebibliography}
\end{document}